\documentstyle[12pt]{article}
\newcommand{\apm}{\!\stackrel{\leftrightarrow\;}{\partial_{\mu}}\!}
\newcommand{\cint}{\int_{\Sigma}d\Sigma}
\title{Is particle creation by the gravitational field consistent 
with energy conservation?}
\author{Hrvoje Nikoli\'c \\
Theoretical Physics Division, Rudjer Bo\v{s}kovi\'{c} Institute, \\
P.O.B. 180, HR-10002 Zagreb, Croatia \\
{\normalsize hrvoje@faust.irb.hr} \\
\makebox[1in]{} \\
}
\date{\today}
\begin{document}
\maketitle
\begin{abstract}
If particle creation 
is described by a Bogoliubov transformation, then, in the 
Heisenberg picture, the raising and lowering operators are 
time dependent. On the other hand, this time dependence is 
not consistent with field equations and the conservation
of the stress-energy tensor.  
Possible physical interpretations and 
resolutions of this inconsistency are discussed.  
\end{abstract}
\vspace*{0.5cm} 
\vspace*{0.9cm}

%\newpage

\noindent
It is widely believed that background gravitational field can 
cause production of particles \cite{bd}. However, the 
theoretical framework that describes this hypothetical effect, 
based on Bogoliubov transformation, is still 
far from being free of fundamental and conceptual problems. 
One of the problems is how particle creation from the vacuum 
can be consistent with the conservation of energy. It is 
often argued that creation of particles causes a backreaction 
on the background metric in such a way that energy is conserved. 
However, this conjecture has never been proved rigorously. 
In this letter we argue that the backreaction cannot solve 
the energy problem if the description of particle creation is based on 
Bogoliubov transformation. More precisely, we show that continuous 
change of the average number of particles is inconsistent with the 
local conservation of the stress-energy tensor. 

The background metric is described by the semi-classical Einstein 
equation
\begin{equation}\label{e1}
\frac{1}{2}g_{\mu\nu}R-R_{\mu\nu}=8\pi G\langle \psi | T_{\mu\nu} 
|\psi \rangle \; .
\end{equation}
We assume that the backreaction caused by all physical processes, including 
a possible particle production, is included in (\ref{e1}). 
We only exclude physical processes related to a collapse of the 
quantum state $|\psi \rangle$, because the semi-classical equation 
is not consistent when the collapse is taken into account \cite{page}.
For simplicity, we assume that (\ref{e1}) refers to quantities 
renormalized such that other possible geometrical terms on the 
left-hand side \cite{bd} are not present.    
Since, by assumption, the backreaction is included exactly, 
equation (\ref{e1}) must be exact. The covariant divergence of the 
left-hand side vanishes identically, so the covariant divergence 
of the right-hand side must vanish too. We work in the Heisenberg picture 
(in which the state $|\psi \rangle$ is constant), which implies that 
the operator equation 
\begin{equation}\label{e2}
D^{\mu}T_{\mu\nu}=0
\end{equation}
must be valid exactly. For simplicity, we assume that 
$T_{\mu\nu}\equiv T_{\mu\nu}(\phi)$ is determined by the action 
of a hermitian scalar field $\phi$ coupled only to the exact  
background metric $g_{\mu\nu}$. Therefore, the corresponding 
curved space-time Klein-Gordon equation 
\begin{equation}\label{e3}
(D^{\mu}\partial_{\mu} +m^2 +\xi R)\phi =0
\end{equation}
should be valid exactly. 

Let $\Sigma(t)$ denote some foliation of space-time into Cauchy 
spacelike hypersurfaces. The coordinates $x=(t,{\bf x})$ are chosen 
such that $t={\rm constant}$ on $\Sigma$. 
We want to show that if 
the average number of particles at $\Sigma$ changes continuously   
with time $t$, then equations (\ref{e2}) and (\ref{e3}) 
are not satisfied. 

The field $\phi$ can be expanded as 
\begin{equation}\label{e4}
\phi(x)=\sum_{k}a_k f_k(x)+ a^{\dagger}_k f^*_k(x) \; .
\end{equation}
Here $f_k(x)$ are solutions of (\ref{e3}) such that they are 
positive-freqeuency solutions at some initial time $t_0$. 
We introduce the vacuum $|0\rangle$ as a state with the 
property $a_k|0\rangle=0$. This state has zero particles at 
$t_0$. A general state $|\psi\rangle$ is constructed by acting 
on $|0\rangle$  
with the operators $a^{\dagger}_k$. To find the 
average number of particles at some other time $t>t_0$, we 
introduce a new set of functions $u_l(x)$ that satisfy 
(\ref{e3}) and are positive-freqeuency solutions at $t$. 
Instead of (\ref{e4}), we can use the expansion 
\begin{equation}\label{e5}
\phi(x)=\sum_{l}A_l u_l(x)+ A^{\dagger}_l u^*_l(x) \; .
\end{equation}
The functions $u_l$ satisfy the relations 
\begin{eqnarray}\label{e9}
& (u_l,u_{l'})=-(u^*_l,u^*_{l'})=\delta_{ll'} \; , & \nonumber \\
& (u_l,u^*_{l'})=(u^*_l,u_{l'})=0 \; , &
\end{eqnarray}
where 
\begin{equation}\label{e8}
(\varphi,\chi)=i\cint^{\mu}\varphi^* \apm \chi \; . 
\end{equation}
When $\varphi(x)$ and $\chi(x)$ are solutions of (\ref{e3}), then 
the scalar product (\ref{e8}) does not depend on $\Sigma$. 
The relations analogous to (\ref{e9})
are also valid for $f_k$.                                           
The two sets of functions are related by the Bogoliubov transformation 
\begin{equation}\label{e6}
u_l(x)=\sum_{k} \alpha_{lk} f_k(x) + \beta_{lk} f^*_k(x) \; ,
\end{equation}
where
\begin{equation}\label{e7}
\alpha_{lk}=(f_k,u_l) \; , \;\;\;\; \beta_{lk}=-(f^*_k,u_l) \; .
\end{equation}
From the requirement that (\ref{e4}) and (\ref{e5}) should be equal, 
one finds
\begin{equation}\label{e10}
A_l=\sum_{k} \alpha^*_{lk} a_k - \beta^*_{lk} a^{\dagger}_k \; .
\end{equation}

Equation (\ref{e10}) is sufficient to express the average number of particles 
in a state $|\psi\rangle$ at the time $t$. However, to see 
the relevance of the stress-energy tensor explicitly, we use 
a formalism which is equivalent, but more complicated. The time 
evolution is generated by the Hamiltonian 
\begin{equation}\label{e11}
H(t)=\int_{\Sigma(t)} d\Sigma^{\mu} n^{\nu} T_{\mu\nu}(\phi) \; ,
\end{equation}
where $n^{\nu}$ is a unit vector normal to $\Sigma(t)$. The time 
dependence of the left-hand side is a consequence of the 
time dependence of the metric $g_{\mu\nu}$. One could also add 
the gravitational contribution to the right-hand side of (\ref{e11}). 
However, this contribution depends only on 
the c-numbers $g_{\mu\nu}$, 
not on the operators $\phi$, so it is physically irrelevant to the time 
evolution of the operators that describe matter. The time 
evolution of the particle-number operator is described by 
\begin{equation}\label{e12}
N(t)=U(t,t_0)N(t_0)U^{\dagger}(t,t_0) \; ,
\end{equation}
where $N(t_0)=\sum_{k} a^{\dagger}_k a_k$. The unitary operator 
$U$ satisfies the Schr\"{o}dinger equation 
\begin{equation}\label{e13}
i\frac{d}{dt}U(t,t_0)=H(t)U(t,t_0) \; .
\end{equation}
Putting (\ref{e5}) into the expression for $T_{\mu\nu}(\phi)$ in 
(\ref{e11}), equations (\ref{e12}) and (\ref{e13}) give \cite{gris}
\begin{equation}\label{e14}
N(t)=\sum_{l} A^{\dagger}_l A_l \; . 
\end{equation}
Of course, we could obtain this directly, without using 
(\ref{e11}), (\ref{e12}) and (\ref{e13}). 

The formalism presented above is not new. However, 
in order to understand the discussion that follows, it was 
important to explicitly explain  
the crucial steps in the derivation of (\ref{e14}). The problem 
is that $A_l$ in (\ref{e5}) are constant operators, so 
(\ref{e14}) implies $dN(t)/dt=0$. In other words, the average number 
of particles does not continuously change with time:
\begin{equation}\label{e15}
\frac{d}{dt}\langle \psi| N(t)|\psi \rangle =0 \; .
\end{equation} 
To avoid this difficulty, one expects that (\ref{e14}) should 
be replaced by an expression of the form 
\begin{equation}\label{e16}
N(t)=\sum_{l} A^{\dagger}_l(t) A_l(t) \; . 
\end{equation}
It is not difficult to understand the origin of this extra 
time dependence. To describe the continuous creation of 
particles, we need a new set of functions $u_l(x)$ for 
{\em each} time $t$. This means that the modes $u_l$ 
possess an extra continuous time dependence, i.e. they become 
functions of the form $u_l(x;t)$. These functions do not 
satisfy (\ref{e3}). However, the functions $u_l(x;\tau)$ 
satisfy (\ref{e3}), provided that $\tau$ is kept fixed 
when the derivative $\partial_{\mu}$ acts on $u_l$. 
For $\varphi(x;t)$, $\chi(x;t)$ being two arbitrary functions with such 
an extra t-dependence, we define the scalar product as 
\begin{equation}\label{e17}
(\varphi,\chi)_t \equiv i\int_{\Sigma(t)} d\Sigma^{\mu} 
\varphi^*(x;\tau) \apm \chi(x;\tau)\; |_{\tau=t} \; , 
\end{equation}
where $\tau$ is kept fixed when the derivative acts. In general, 
this scalar product depends on $t$, even if  
$\varphi(x;\tau)$ and $\chi(x;\tau)$ satisfy (\ref{e3}) when  
$\tau$ is kept fixed. The functions $u_l(x;t)$ satisfy 
relations (\ref{e9}) with respect to the scalar product 
(\ref{e17}). However, the Bogoliubov coefficients (\ref{e7}), 
calculated using the scalar product (\ref{e17}), become 
$t$-dependent. Therefore, (\ref{e10}) should be replaced by 
\begin{equation}\label{e18}
A_l(t)=\sum_{k} \alpha^*_{lk}(t) a_k - \beta^*_{lk}(t) a^{\dagger}_k \; .
\end{equation}
This relation can be derived from the requirement that (\ref{e4})  
should be equal to 
\begin{equation}\label{e19}
\phi(x)=\sum_{l}A_l(t) u_l(x;t)+ A^{\dagger}_l(t) u^*_l(x;t) \; ,
\end{equation}
or 
\begin{equation}\label{e20}
\phi(x)=\sum_{l}A_l(t) u_l(x)+ A^{\dagger}_l(t) u^*_l(x) \; ,
\end{equation}
provided that the $t$-dependence of $A_l(t)$ is treated as an 
extra $t$-dependence in (\ref{e17}). In other words, (\ref{e18}) 
is valid if the relation 
$(u_{l'},A_l(t)u_l)_t = A_l(t)\, (u_{l'},u_l)_t$ (as well as 
other similar relations) is valid. 

Now it seems that we have a consistent derivation of the extra 
time dependence in (\ref{e16}). However, it is not consistent. 
The fields (\ref{e19}) and (\ref{e20}) do {\em not} satisfy the 
Klein-Gordon equation (\ref{e3}). Therefore, 
(\ref{e19}) and (\ref{e20}) cannot be equal to (\ref{e4}). 
This implies that equation (\ref{e18}), which expresses the 
equality of (\ref{e4}), (\ref{e19}) and (\ref{e20}),  
cannot be consistent either. This is, indeed, true because the scalar 
product (\ref{e17}) is inconsistently defined. Namely, the 
``extra" t-dependence cannot really be distinguished from the 
``regular" t-dependence, because one can always write 
$\varphi(x;t)\equiv\tilde{\varphi}(x)$. On the other hand, 
to obtain a relation similar to (\ref{e16}) by using 
(\ref{e11}), (\ref{e12}) and (\ref{e13}), one needs to put 
(\ref{e19}) or (\ref{e20}) into the expression for 
$T_{\mu\nu}(\phi)$. Since (\ref{e19}) and (\ref{e20}) do not 
satisfy (\ref{e3}), this $T_{\mu\nu}$ does not satisfy (\ref{e2}). 
Since the expression for $T_{\mu\nu}(\phi)$ involves the 
time derivatives $\dot{\phi}$, the resulting expression for 
$N(t)$ will take the form 
\begin{equation}\label{e21}
N(t)=\sum_{l} A^{\dagger}_l(t) A_l(t) 
+ {\cal N}(\dot{A}_l(t), \dot{u}_l(\tau,{\bf x};t)|_{\tau=t}) . 
\end{equation}
The extra term ${\cal N}$ obtained using (\ref{e19}) is not the same  
as that obtained using (\ref{e20}). 
This term is negligible if we assume that the 
change of the average number of particles is slow, i.e. that 
$\dot{A}_l(t)\approx 0$, $\dot{u}_l(\tau,{\bf x};t)|_{\tau=t}\approx 0$. 
If we take this approximation, then (\ref{e3}) and (\ref{e2}) are 
approximately valid. However, although the particle production 
can be slow, the total number of produced particles 
can be significant after a long time. 
Similarly, the total produced energy can be
significant after a long time.  
    
Our results can be summarized as follows: If particle creation 
is described by a Bogoliubov transformation, then, in the 
Heisenberg picture, the raising and lowering operators are 
time dependent. On the other hand, this time dependence is 
not consistent with the field equations and the conservation 
of the stress-energy tensor. 
Below we discuss several possible approaches 
to the resolution of this problem and show that none of them is  
completely satisfactory. 

One possibility is to define the average of the stress-energy tensor 
in an independent way, without using (\ref{e19}) or (\ref{e20}). 
Indeed, the stress-energy tensor is often defined  
using a formalism based on the 
Schwinger-DeWitt representation of the Green function \cite{bd}. The 
stress-energy tensor defined in this way is automatically 
conserved, but 
is {\em not} represented as $\langle\psi | T_{\mu\nu}|\psi \rangle$ for some 
operator $T_{\mu\nu}$. Such an approach leads to an argument 
that particle production is consistent with the local 
conservation of the stress-energy tensor \cite{bd}. However, 
we do not find this argument satisfactory, 
because the formalism 
in which the average number of particles is described by an operator, 
whereas the average stress-energy tensor is not described by an operator, 
does not seem to be consistent. Moreover, if one does not define 
the operator $T_{\mu\nu}$, then one cannot describe the particle 
production by the formalism based on (\ref{e11}). Finally, 
it is not clear in such an approach whether the Klein-Gordon 
equation (\ref{e3}) is satisfied.  

Another possibility (that can be combined with the possibility 
above) is that a well-defined operator $N(t)$ simply 
does not exist, even when the foliation $\Sigma(t)$ is chosen. 
Instead, the number of particles should be defined operationally, 
by the response of a ``particle" detector of the 
Unruh-DeWitt type \cite{unruh,dewitt}. Such detectors need a 
long time to measure the number of particles in a reliable way.           
The average number of particles at a given time $t$ loses its 
meaning. However, since all other observables in quantum mechanics 
can be represented by well-defined hermitian operators that 
evolve continuously with time and  
do not require a model of the correspoding detector, it is not 
clear why the number of particles should be an exception. 

At this point it is instructive to compare the concepts of 
energy and particle number in flat space-time. In this case,  
both quantities are conserved. On the other hand, both quantities 
obey certain approximate uncertainty relations \cite{park}    
\begin{eqnarray}\label{parker}
& \Delta E \Delta t \geq 1 \; , & \nonumber \\
& \Delta N \Delta t \geq  m^{-1} \; , &
\end{eqnarray}
where $m$ is the mass of the particle and $\Delta t$ is the time interval 
during which the measurement is performed. These relations cannot 
be derived from some fundamental quantum principles. They merely 
express the uncertainties related to typical methods of measurement. 
Actually, it is, in principle, possible to measure energy with an arbitrary 
accuracy inside an arbitrarily small time interval \cite{ahar}. There is no 
reason why this would not be the case for the number of particles as well. 
The uncertainty relations do not imply that the operators $H(t)$ and $N(t)$ 
are not well defined, even when they are not conserved.         

Our discussion suggests a new approach to the resolution of the 
problem of inconsistency between particle creation and energy 
conservation. In this approach, equation (\ref{e15}) is interpreted 
as conservation of the particle number when the 
classical gravitational interaction 
described by (\ref{e3}) is the only interaction. It seems that in such an 
approach one has to reject a common belief that the definition 
of particles should be closely related to the definition of positive 
frequencies. We shall study such a possibility in more detail 
elsewhere.    

A close relationship between the non-conservation of energy and 
particle number can also be seen in the following way: Assume that 
space-time is flat at $t_0$ and $t$, but not at the intermediate 
times. In this case, the non-conservation of energy is obvious 
from the relations
\begin{eqnarray}\label{flaten}
& H(t_0)=\sum_{k} \omega_k (a^{\dagger}_k a_k +\frac{1}{2}) \; , & 
 \nonumber \\
& H(t)=\sum_{k} \omega_k (A^{\dagger}_k A_k +\frac{1}{2}) \; . &
\end{eqnarray}
For example, $\langle 0|H(t)|0\rangle -\langle 0|H(t_0)|0\rangle 
=\sum_{k}\sum_{k'} \omega_k |\beta_{kk'}|^2/2$. The fact that the 
energy should be conserved suggests that certain operators 
should be renormalized such that the renormalized Hamiltonian is 
conserved. A formalism which renormalizes the Hamiltonian  
should also renormalize the number of particles. The equations 
given above 
suggest the following renormalization of the raising and lowering 
operators: 
\begin{equation}\label{ren}
A^{{\rm ren}}_k =A_k-\sum_{k'}[ (\alpha^*_{kk'}-\delta_{kk'})a_{k'}
-\beta^*_{kk'}a^{\dagger}_{k'} ] \; .
\end{equation}
Since $A^{{\rm ren}}_k=a_k$, such a renormalization leads to 
$H^{{\rm ren}}(t)=H(t_0)$. A necessary consequence of such a
renormalization is that there is no particle production because 
$N^{{\rm ren}}(t)=N(t_0)$. The problem with this heuristic 
argument against particle production is that (\ref{ren}) is 
obtained in an ad hoc way.  
One needs a derivation that starts from more fundamental 
principles.   

There is no doubt that the usual concept of particles that emerges 
from free fields in Minkowski space-time should be modified 
significantly when the generalization to arbitrary space-time 
is considered. Our analysis demonstrates that existing 
achievements in this field are far from being completely 
satisfying. 
Further investigation is needed in order to 
formulate a closed and consistent theory. 
 
\vspace{0.5cm}
\noindent
{\bf Acknowledgement}
\vspace{0.5cm}

This work was supported by the Ministry of Science and Technology of the
Republic of Croatia under Contract No. 00980102.

\end{document}